\documentclass[twocolumn,amsmath,amssymb]{revtex4}

\usepackage{amssymb}

\usepackage{graphicx}
\usepackage{dcolumn}
\usepackage{bm}
\newcommand{\etal}{\emph{et al.}}
\newcommand{\be}{\begin{equation}}
\newcommand{\ee}{\end{equation}}
\newcommand{\bfig}{\begin{figure}}
\newcommand{\efig}{\end{figure}}

\begin{document}




\title{Thermal-Hall conductivity and long-lived quasiparticles in CeCoIn$_5$
} 
\author{Y. Onose$^1$, N. P. Ong$^1$ and C. Petrovic$^2$\footnote{Proceedings 
of M$^2$S-HTSC-VIII, Dresden 2006, Physica C, \emph{in press}}
}
\affiliation{
$^1$Department of Physics, Princeton University, Princeton, NJ 08544, USA\\
$^2$Department of Physics, Brookhaven National Laboratory, Upton, N.Y. 11973, USA
}

\begin{abstract}
In CeCoIn$_5$, the thermal conductivity $\kappa_{xx}$ and Hall conductivity $\kappa_{xy}$ below $T_c$
display large anomalies below $T_c$.  The strong suppression of the anomalies in weak fields implies the 
existence of long-lived quasiparticles.  We also discuss briefly the Wiedemann-Franz ratio and the existence of
a strongly field-dependent spin-fluctuation heat current.

\end{abstract}

\maketitle                   


The heavy fermion system CeCoIn$_5$ displays several interesting properties
in its superconducting state.  These include the FFLO state in an in-plane 
magnetic field $\bf H$~\cite{FFLO}, a large Nernst signal observed at temperatures
$T<$ 30 K~\cite{Nernst}, an in-plane resistivity that is $T$-linear
below 20 K, and a Hall coefficient that is strongly $T$ dependent~\cite{Matsuda1}.
Recently, Kasahara \etal~\cite{Kasahara} measured the thermal conductivity $\kappa_{xx}(T,H) = \kappa(T,H)$ 
and thermal Hall conductivity $\kappa_{xy}$ and inferred a long quasiparticle (qp) 
mean-free-path $\ell$ below $T_c$ = 2.2 K.

\begin{figure}[h]			
\includegraphics[width=7cm]{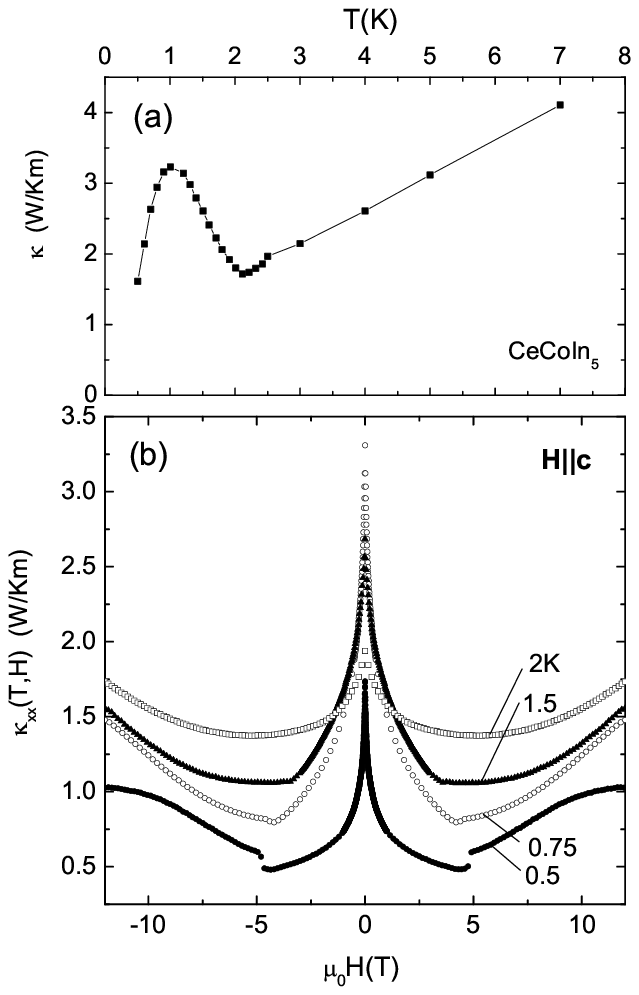}
\caption{\label{kxx}  (Panel a) The $T$ dependence of the in-plane thermal conductivity $\kappa$
in $H$ = 0 in CeCoIn$_5$.  The anomaly below $T_c$ is 4--5 times larger than in earlier reports.
(Panel b)  The field dependence of $\kappa$ at selected $T$ below $T_c$.  With decreasing $T$,
the zero-field anomaly rises to a sharp, narrow peak.  At $H_{c2}$, $\kappa$ displays a kink or step.
}
\end{figure}
We report measurements of $\kappa_{ij}$ in crystals of CeCoIn$_5$ with exceptionally long $\ell$.
Figure \ref{kxx}a shows $\kappa(T,0)$ at $H$=0.  Below $T_c$, $\kappa(T,0)$
rises steeply to a prominent peak at 1 K, reminiscent of the peak in high-purity YBCO~\cite{Zhang01}.
As shown in Fig. \ref{kxx}b, the peak anomaly -- 4-5 times larger than in Ref.~\cite{Kasahara} -- is extremely 
sensitive to $\bf H||c$.  Above $T_c$, the curve of $\kappa$ displays moderately strong field dependence.  Below
$T_c$, however, a sharp quasiparticle peak develops at $H=0$ and rises rapidly.  The narrow 
spike in $\kappa$ which reflects the strong suppression of the qp heat
current in weak fields.  The narrow spike is absent in earlier experiments~\cite{Kasahara}.  
In the normal state above the upper critical field $H_{c2}$ (indicated by the
step increase), $\kappa$ remains strongly $H$ dependent.

The thermal Hall conductivity $\kappa_{xy}$ detects the qp heat current of alone.
Above 20 K, $\kappa_{xy}$ is nearly $H$-linear to 12 T, as expected of weak-field 
Hall response.  As $T$ falls towards $T_c$, strong curvature becomes evident 
below 1 T, while the initial Hall slope increases sharply.  Between $T_c$ and 0.5 K, 
a new anomaly appears in weak $H$ which is the Hall analog of the narrow spike 
in $\kappa$.  In weak $H$, $\kappa_{xy}$ rises very steeply
to a peak centered at 0.1--0.2 T before falling to a ``plateau'' value.  
Above $H_{c2}$, $\kappa_{xy}$ increases steeply once more to large values.

\begin{figure}[h]		
\includegraphics[width=8cm]{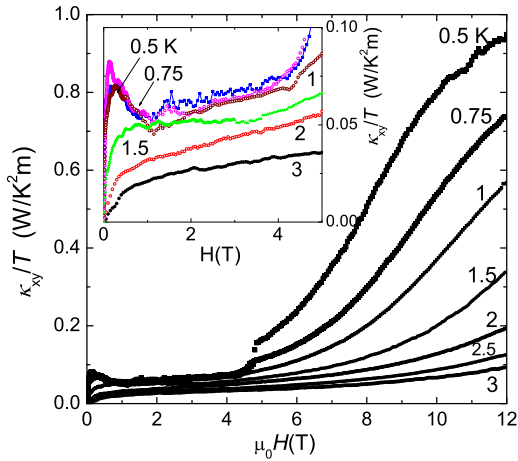}
\caption{\label{kxy}  Curves of the thermal Hall conductivity (divided by $T$) $\kappa_{xy}/T$ vs. $H$.  
Below 1 K, $\kappa_{xy}/T$ saturates to a $T$-independent profile with a
prominent anomaly in weak $H$ (inset).  At low $T$, $\kappa_{xy}/T$ sharply increases 
above $H_{c2}$ (step feature).
}
\end{figure}

The unusual behavior of $\kappa$ and $\kappa_{xy}$ reveal several interesting 
low-$T$ features in both the vortex state and the state above $H_{c2}$ at low $T$.   
Below $T_c$, the strong sensitivity of $\kappa$ and $\kappa_{xy}$ to weak $H$ shows that
the large thermal anomaly shown in Fig. \ref{kxx}a is entirely electronic in origin.  It reflects the
steep increase in $\ell$ below $T_c$.  Moreover, the persistence of a large $\kappa_{xy}$
far below $T_c$ requires a sizeable qp population, which implies the existence of line nodes
on the Fermi Surface.  This was previously inferred from the 4-fold variation of $\kappa$
and the heat capacity $c_p$ with field angle~\cite{Matsuda2,heatcap} in crystals with higher degree of disorder.

The curves of $\kappa_{xy}/T$ below $T_c$ (Fig. \ref{kxy}) show that the qp behavior is qualitatively
distinct in the vortex-solid state below $H_{c2}\sim$ 5 T, and in the normal state above.  
In the former, $\kappa_{xy}/T$ assumes a profile that is $T$ independent below 1 K.  
Interestingly, at the lowest temperatures (0.5--1 K) the value of $\kappa_{xy}/T$ is 
independent of $T$ to within the experimental uncertainty.  After going through the weak-field
peak, the curve settles into a nearly flat portion that extends to 4 T, before rising vertically at
$H_{c2}$.  By contrast, the strong field variation of $\kappa_{xy}/T$ above $H_{c2}$ is striking. In 
particular, the curvature just above $H_{c2}$ is quite pronounced.  These features imply
that the scattering rate $\Gamma$ is strongly suppressed by $H$, resulting in a significant
increase in $\ell$.

From the measured Hall conductivity $\sigma_{xy}$ and $\kappa_{xy}$ we may find the Wiedemann-Franz 
ratio~\cite{Zhang00}.  The Lorenz number $L$ is found to lie within 10$\%$ of the Sommerfeld value 
$L = \pi^2/3(k_B/e)^2$.  Knowing $L$ at each $T$ above $T_c$, 
we infer the qp thermal conductivity $\kappa_e(T,H)$ from the measured conductivity $\sigma = \sigma_{xx}$.
Subtracting $\kappa_e(T,H)$ from the observed $\kappa(T,H)$, we find a remaining term  
that is strongly $H$ dependent even high above $T_c$.  We identify this term with the thermal conductivity $\kappa_s(T,H)$ 
due to spin fluctuations (the phonon conductivity $\kappa_{ph}(T)$ is $H$ independent).
At low $T$, $\kappa_s(T,H)$ accounts for most of the non-qp heat current.

Research at Princeton and the Brookhaven National Laboratory was 
supported, respectively, by U.S. NSF (DMR 0213706) and Department of Energy (DE-Ac02-98CH10886).



\end{document}